\documentclass[prd,12pt,showpacs,preprintnumbers,floatfix,amsmath,amssymb,nofootinbib,tightenlines]{revtex4}
\usepackage{graphicx}

\newcommand{\beqn}{\begin{eqnarray}}
\newcommand{\eeqn}{\end{eqnarray}}

\begin{document}

\title{Bulk viscosity of hot dense Quark matter in PNJL model}
\author{Shi-Song Xiao }
\affiliation{Department of Computer Science , Central China Normal University, Wuhan, 430079, China}

\author{Pan-pan Guo }
\affiliation{Institute of Particle Physics, Central China  Normal
University, Wuhan 430079, China}
\author{Le Zhang }
\affiliation{Institute of Particle Physics, Central China  Normal
University, Wuhan 430079, China}

\author{De-fu Hou}
\affiliation{Institute of Particle Physics, Central China Normal
University, Wuhan 430079, China}

\begin{abstract}

 Starting from the Kubo formula  and the the QCD low energy theorem, we study the  the
bulk viscosity of hot dense quark matter  in the PNJL model  from the  equation of state .    We show that the
 bulk
viscosity has a sharp peak near the chiral phase transition, and the ratio
of bulk viscosity over entropy  rises dramatically
in the vicinity of the  phase transition. These results agrees with that from lattice and other model calculations.  In addition, we show that  the increase of chemical
potential raises the bulk viscosity.

\end{abstract}


\maketitle

The study of transport properties of strong interacting matter has been attracting many interests. It is very important for hydrodynamic simulations of heavy-ion collisions and for understanding properties of compact stars\cite{r1,r2,r3,r4,HC}.
. Shear  viscosity $\eta $ characterizes how fast a system goes back to equilibrium under a shear mode perturbation . It is  believe that the quark{gluon plasma (QGP) found in relativistic heavy-ion collider (RHIC) is strongly coupled, which is contrast to the weak coupling picture expected earlier. This is the so-called sQGP.   Lattice Monte Carlo simulation4 on sQGP demonstrated that the ratio of the shear viscosity to the entropy
density is rather small but still probably larger than the universal lower bound $1/{4\pi}$ which is obtained from Ads/CFT duality\cite{sv1}. The experimental extracted value with viscous hydrodynamics combining with a microscopic transport model lies within in the range $1\sim 2.5$ times of the lower bound\cite{HC}.

Bulk viscosity describes how fast a system goes back to equilibrium under  a uniform expansion,relating to the deviation from the conformal invariance   of the system . It vanishes when the system has a conformal equation of state, therefore the sharp peak of the bulk viscosity would strongly affect the physics of the QCD matter near critical temperature and is very important for the study of QCD phase structure. Also bulk  viscosity  affects the Elliptic flow near QCD phase transition in the Relativistic heavy ion collisions\cite{r5,r6}.  The study of bulk viscosity is also  important for the physics  of  compact stars\cite{r1,r2,r3,r4}.

Recently lattice QCD calculation shows  that the trace of energy-momentum tensor anomaly   and  the ratio of the bulk viscosity $\zeta$ over
entropy density $s$ have  a sharp peak  or diverge near phase transition\cite{r14, r15,bulk1,bulk2}. Such a sharp peak behavior of $\zeta$ has also
been observed in many model calculations \cite{bulk3,bulk4,bulk5,bulk9}.


At present most of the calculations are for  zero  baryon
density\cite{bulk1,r9} , except a few papers trying to estimate the bulk viscosity with finite  density\cite{jiang,bulk7}.For example, in Ref. \cite{bulk7} the authors study the viscosity at
finite $\mu$ with Nambu-Jona-Lasinio (NJL) model. In Ref. \cite{bulk8} the authors study the viscosity of strange quark matter  at finite $\mu$ with quasi particle model. While the bulk viscosity was studied in \cite{jiang} with Dyson-Schwinger equations at finite $\mu$ but zero temperature. Here we promote the calculation of bulk viscosity  to both finite temperature and finite  baryon density in PNJL model incorporating both confinement and chiral symmetry  in this paper.

The bulk viscosity of hot dense quark matter is related to the
retarded Green's function of the  trace of the
energy-momentum tensor by  Kubo formula.   Using low energy theorems
at finite temperature and chemical potential,we can extract the bulk
viscosity of hot dense quark matter from the small frequency ansatz.

From Kubo  formula,we can express the bulk viscosity at Lehmann
representation\cite{r13}
\begin{equation}
\zeta=\frac{1}{9}\lim_{\omega\rightarrow0}\frac{1}{\omega}\int_{0}^{\infty}dt\int
d^{3}\overrightarrow{r}\exp{(i\omega
t)}\langle[\theta_{\mu}^{\mu}(x),\theta_{\mu}^{\mu}(0)]\rangle.\\\label{1}
\end{equation}
Where $\omega$ is the frequency, $\theta_{\mu}^{\mu}$ is the trace
of the energy-momentum tensor.Using Fourier transform and
P-invariance,the formula is changed as
\begin{eqnarray}
\nonumber\zeta &=&\frac{1}{9}\lim_{\omega\rightarrow0}\frac{1}{\omega}\int_{0}^{\infty}dt\int d^{3}\overrightarrow{r}\exp{(i\omega t)}iG^{R}(x)\\
\nonumber&=&\frac{1}{9}\lim_{\omega\rightarrow0}\frac{1}{\omega}iG^{R}(\omega,\overrightarrow{0})\\&=&-\frac{1}{9}\lim_{\omega\rightarrow0}\frac{1}{\omega}ImG^{R}(\omega,\overrightarrow{0}).
\label{2}
\end{eqnarray}
In Lehmann representation,the Green'function is related to spectral
density $\rho(\omega,\overrightarrow{p})=-\frac{1}{\pi}ImG^{R}(\omega,\overrightarrow{p}).$
For Kramers-Kroning relation,we can obtain
\begin{eqnarray}
\nonumber
G^{R}(\omega,\overrightarrow{p})&=&\frac{1}{\pi}\int_{-\infty}^{\infty}\frac{ImG^{R}(u,\overrightarrow{p})}{u-\omega-i\varepsilon}du
\\&=&\int_{-\infty}^{\infty}\frac{\rho(u,\overrightarrow{p})}{\omega-u+i\varepsilon}du.\label{3}
\end{eqnarray}
The Euclidean Green'function is
\begin{equation}
\nonumber
G^{E}(\omega,\overrightarrow{p})=-G^{R}(i\omega,\overrightarrow{p}),\omega>0\
\end{equation}
Using the formula(3)we have
\begin{equation}
G^{E}(0,\overrightarrow{0})=2\int_{0}^{\infty}\frac{\rho(u,\overrightarrow{0})}{u}du.\label{4}
\end{equation}

For QCD, the trace of energy-momentum stress tensor reads

\begin{equation}
\theta ^\mu_\mu=m_q\bar{q}q+\frac{\beta(g)}{2g}
F_{\mu\nu}^aF^{a\mu\nu}
\equiv\theta_F+\theta_G\label{trace anomaly},
\end{equation}
where $g$ is the strong coupling constant, $\theta_F$ and $\theta _G$ are the contribution of quark fields and of gluon field, respectively, and $\beta(g)$ is the QCD
$\beta$-function which determines the running behavior of $g$. In
Eq. (\ref{trace anomaly}) $q$ are quark fields with two flavors (in
this letter we will limit ourselves in two flavor case and set the
current quark mass $m_u=m_d=m$).

 From the QCD low-energy theorems at
finite temperature $T$ and $\mu$ \cite{LET1}, one can find

\begin{equation}
\left(T\frac{\partial}{\partial
T}+\mu\frac{\partial}{\partial\mu}-d\right)
\langle\hat{\mathcal {O}}\rangle_T=\int d^4x \langle T_t\{\theta_G(x), \hat{\mathcal{O}}(0)\}\rangle,
\end{equation}

where $d$ is the canonical dimension of the operator  $\hat{\mathcal{O}}$. Using the above equation, one has

\begin{equation}
\left(T\frac{\partial}{\partial
T}+\mu\frac{\partial}{\partial\mu}-4\right)\langle
\theta _G\rangle_T
=\int d^4x\langle T_t
\{\theta_G(x), \theta_G(0)\}\rangle,
\end{equation}

\begin{equation}
\left(T\frac{\partial}{\partial
T}+\mu\frac{\partial}{\partial\mu}-3\right)\langle
\theta_F\rangle_T
=\int d^4x\langle T_t
\{\theta_G(x),\theta_F(0)\}\rangle.
\end{equation}

From the above two relations one obtains
\begin{eqnarray}\label{bulkvis2}
&&9\zeta\omega_0=\int d^4x\langle\,T_t\{\theta_\mu^\mu(x),
\theta_\mu^\mu(0)\}\rangle\nonumber\\
&=&\left(T\frac{\partial}{\partial
T}+\mu\frac{\partial}{\partial\mu}-4\right)\langle
\theta_G\rangle_T+2\left(T\frac{\partial}{\partial
T}+\mu\frac{\partial}{\partial\mu}-3\right)\langle
\theta_F\rangle_T \nonumber \\
&& +\int d^4x\langle\,T_t\{\theta_F(x)
\theta_F(0)\}\rangle\nonumber\\
&\approx& \left(T\frac{\partial}{\partial
T}+\mu\frac{\partial}{\partial\mu}-4\right)
\langle\theta _\mu^\mu\rangle_T+\left(T\frac{\partial}{\partial
T}+\mu\frac{\partial}{\partial\mu}
-2\right)\langle\theta_F\rangle_T \nonumber\\
&=&f_1(T,\mu)(\varepsilon-3P)  +f_2(T,\mu) \langle\theta_F\rangle_T,
\end{eqnarray}
where
\begin{eqnarray}
 f_1(T,\mu)&=&\left(T\frac{\partial}{\partial
T}+\mu\frac{\partial}{\partial\mu}-4\right)  \nonumber\\
f_2(T, \mu)&=&\left(T\frac{\partial}{\partial
T}+\mu\frac{\partial}{\partial\mu}-2\right),
 \end{eqnarray}
 and $\varepsilon$ is the energy density and $P$ is the
pressure density of QCD. Here, because the current quark mass $m$ of u and d quark is very small, in deriving Eq. (\ref{bulkvis2}) we have neglected the term proportional to $m^2$.

The low energy theorems adapt to long distance,low frequency and
strong coupling QCD\cite{r11}\cite{r12}.Using the non-perturbation
theory,the Euclidean Green's function can be represented as
\begin{eqnarray}
\nonumber G^{E}(0,\overrightarrow{0})&=&\int d^{4}x<T\theta(x),\theta(0)>\\
&=&f_1(T,\mu) <\theta> _{T} +f_2(T,\mu) \langle\theta_F\rangle_T. \label{5}
\end{eqnarray}
Where $<\theta> _{T}$ is the trace of the energy-momentum tensor.Its
average value in zero temperature is $<\theta>
_{0}=-4|\varepsilon_{v}|$, $\varepsilon_{v}$ is the vacuum energy
density,including the quark condensates and the gluon condensates in
our work.In the low energy theorems,the difference of energy density
and the pressure corresponds to non-zero vacuum expectation value of
the energy-momentum tensor $\varepsilon-3P=<\theta> _{T}-<\theta>
_{0}$. Analogously,  $\langle\theta_F\rangle_T=<m \bar q q>_T + < m \bar q q>_0$ . Using the PCAC relations, we can express the vacuum expectation value $<m \bar q q>_0$ through the Pion and Kaon masses and
decay constants $ < m \bar q q>_0=- M^2_{\pi} f^2_{\pi} -M^2_k f^2_k$. Using these relations ,combining the formula(4)and (5),we
obtain\cite{bulk1}:
\begin{eqnarray}
\nonumber &&2\int_{0}^{\infty}\frac{\rho(u,\overrightarrow{0})}{u}du=({T\frac{\partial}{\partial T}+\mu\frac{\partial}{\partial\mu}-4})<\theta> _{T}\nonumber\\
&+&\left(T\frac{\partial}{\partial
T}+\mu\frac{\partial}{\partial\mu}
-2\right)\langle\theta_F\rangle_T\nonumber\\
&=& (T\frac{\partial}{\partial \nonumber
T}+\mu\frac{\partial}{\partial\mu}-4)(\varepsilon-3P-4|\varepsilon_{v}|+ < m \bar q q>_0 )
\nonumber\\
&+&\left(T\frac{\partial}{\partial
T}+\mu\frac{\partial}{\partial\mu}
-2\right)(<m \bar q q>_T + < m \bar q q>_0). \label{6}
\end{eqnarray}
This formula don't include the perturbative contribution as long as
we consider the strong coupling situation. So we can use the
following ansatz in the small frequency region\cite{bulk1}
\begin{equation}
\nonumber\frac{\rho(\omega,\overrightarrow{0})}{\omega}=\frac{9\zeta\omega^{2}}
{\pi\left({\omega_{0}^{2}+\omega^{2}}\right)} \\
\end{equation}
Where $\zeta$ is the bulk viscosity and $\omega_{0}$ is a scale at
which the perturbation theory becomes valid, $\omega_{0}\sim T$.
Using this ansatz and the formula(6),we extract the bulk viscosity:
\begin{eqnarray}
&&\zeta=\frac{1}{9\omega_{0}}(T\frac{\partial}{\partial
T}+\mu\frac{\partial}{\partial\mu}-4)(\varepsilon-3P-4|\varepsilon_{v}|+ < m \bar q q>_0)\, \label{7}\\
&&+\frac{1}{9\omega_{0}}\left(T\frac{\partial}{\partial
T}+\mu\frac{\partial}{\partial\mu}
-2\right)(<m \bar q q>_T + < m \bar q q>_0)\nonumber
\end{eqnarray}

The NJL(Nambu-Jona-Lasino)model is based on an effective lagrangian
of relativistic fermions which interact through local
current-current couplings.It can illustrate the transmutation of
originally light quarks into massive quasi-particles,and the
spontaneously broken chiral symmetry.But the quark confinement is
missing in the NJL model.  The de-confinement phase transition is
characterized by spontaneous breaking of the Z(3)center symmetry of
QCD.The corresponding order parameter is the Polyakov loop (p-loop). So the
PNJL model introduce both the chiral condensate
$\langle\overline{\Psi}\Psi\rangle$ and the p-loop $\Phi$ coupling
to the quarks to solve the problem of the NJL model\cite{r10,PNJL}.

The PNJL model is an effective method to deal with the
non-perturbative QCD. So the bulk viscosity extracted from the
formula in the low energy theorems can be calculated in this model.
The Lagrangian of two-flavor PNJL model at finite chemical potential
is given by\cite{PNJL}
\begin{eqnarray}
\nonumber\mathcal{L}_{PNJL}&=&\overline{q}\left({i\gamma^{\mu}D_{\mu}-\widehat{m}}\right)q+g\Big[\left({\overline{q}q}\right)^{2}+\left({\overline{q}i\gamma_{5}\overrightarrow{\tau}q}\right)^{2}\Big]\\
&-& \mathcal{U}(\Phi(A),\overline{\Phi}(A),T)\label{8}
\end{eqnarray}
Where $D^{\mu}=\partial^{\mu}-iA^{\mu},A^{\mu}=\delta_{\mu0}A^{0}$.
The effective potential $\mathcal{U}$ is expressed in terms of the
traced p-loop $\Phi=\frac{Tr_{c}L}{N_{C}}$ and its conjugate
$\overline{\Phi}=\frac{Tr_{c}L^{\dag}}{N_{C}}$,where
$L=\exp(\frac{iA_{4}}{T})$,$A_{4}$ is the gauge field.
\begin{eqnarray}
\nonumber\frac{\mathcal{U}(\Phi,\overline{\Phi},T)}{T^{4}}&=&-\frac{b_{2}(T)}{2}\overline{\Phi}\Phi-\frac{b_{3}}{6}\left({\overline{\Phi}^{3}+\Phi^{3}}\right)+\frac{b_{4}}{4}\left({\overline{\Phi}\Phi}\right)^{2};\\
\nonumber
b_{2}(T)&=&a_{0}+a_{1}\frac{T_{0}}{T}+a_{2}\left({\frac{T_{0}}{T}}\right)^{2}+a_{3}\left({\frac{T_{0}}{T}}\right)^{3}.
\end{eqnarray}
The parameters in the effective potential are chosen in the
following Table\cite{r10}.
\begin{center}
\begin{ruledtabular}
\begin{tabular}{cccccc}
$a_{0}$&$a_{1}$&$a_{2}$&$a_{3}$&$b_{3}$&$b_{4}$\\
\hline
6.75&-1.95&2.625&-7.44&0.75&7.5\\
\end{tabular}
\end{ruledtabular}
\end{center}
With the definition of the chiral condensate
$\sigma=\langle\overline{q}q\rangle$ and the constituent quark mass
$M=m-2g\sigma$ the grand potential density is given by
\begin{eqnarray}
\nonumber\Omega(\Phi,\overline{\Phi},M,T,\mu)&=&\mathcal{U}(\Phi,\overline{\Phi},T)+g\langle\overline{q}q\rangle^{2}\\
\nonumber&-&2N_{C}N_{f}\int\frac{d^{3}p}{(2\pi)^{3}}E_{p}\\
\nonumber&+&2N_{f}T\int\frac{d^{3}p}{(2\pi)^{3}}[\ln{N_{\Phi}^{+}(E_{p})}\\
&+&\ln{N_{\Phi}^{-}(E_{p})}]\label{9}.
\end{eqnarray}
Where
\begin{eqnarray}
\nonumber\frac{1}{N_{\Phi}^{+}(E_{p})}&=&1+3(\Phi+\overline{\Phi}\exp{(-\beta E_{p}^{+})})\exp{(-\beta E_{p}^{+})}\\
\nonumber&+&\exp{(-3\beta E_{p}^{+})}\\
\nonumber\frac{1}{N_{\Phi}^{-}(E_{p})}&=&1+3(\overline{\Phi}+\Phi\exp{(-\beta E_{p}^{-})})\exp{(-\beta E_{p}^{-})}\\
\nonumber&+&\exp{(-3\beta E_{p}^{-})}
\end{eqnarray}
$E_{p}=\sqrt{p^{2}+M^{2}}$ is the quasi-particle energy for the
quarks.$E_{p}^{\pm}=E_{p}\mp\mu$,$\mu$ is the quark chemical
potential.Here we consider the isospin symmetry. Now we introduce
the mean-field approach by minimizing $\Omega$ with respect to
$\sigma,\Phi$ and $\overline{\Phi}$, the mean-field equations is
given by
\begin{eqnarray}
\nonumber\sigma&=&-6N_{f}\int\frac{d^{3}p}{(2\pi)^{3}}E_{p}\frac{M}{E_{p}}[\theta(\Lambda^{2}-p^{2})\\
&-&M_{\Phi}^{+}(E_{p})N_{\Phi}^{+}(E_{p})-M_{\Phi}^{-}(E_{p})N_{\Phi}^{-}(E_{p})];\label{10}\\
\nonumber0&=&\frac{T^{4}}{2}[-b_{2}(T)\overline{\Phi}-b_{3}\Phi^{2}+b_{4}\Phi\overline{\Phi}^{2}]\\
\nonumber&-&12T\int\frac{d^{3}p}{(2\pi)^{3}}[\exp{(-2\beta E_{p}^{+})}N_{\Phi}^{+}(E_{p})\\
&+&\exp{(-\beta E_{p}^{-})}N_{\Phi}^{-}(E_{p})];\label{11}\\
\nonumber0&=&\frac{T^{4}}{2}[-b_{2}(T)\Phi-b_{3}\overline{\Phi}^{2}+b_{4}\overline{\Phi}\Phi^{2}]\\
\nonumber&-&12T\int\frac{d^{3}p}{(2\pi)^{3}}[\exp{(-\beta E_{p}^{+})}N_{\Phi}^{+}(E_{p})\\
&+&\exp{(-2\beta E_{p}^{-})}N_{\Phi}^{-}(E_{p})];\label{12}
\end{eqnarray}
The limits of integration is $0\sim\Lambda$ which is a global
cutoff\cite{r10}.Where
\begin{eqnarray}
\nonumber M_{\Phi}^{+}(E_{p})&=&(\Phi+2\overline{\Phi}\exp{(-\beta E_{p}^{+})})\exp{(-\beta E_{p}^{+})}\\
\nonumber&+&\exp{(-3\beta E_{p}^{+})},\\
\nonumber M_{\Phi}^{-}(E_{p})&=&(\overline{\Phi}+2\Phi\exp{(-\beta E_{p}^{-})})\exp{(-\beta E_{p}^{-})}\\
\nonumber&+&\exp{(-3\beta E_{p}^{-})}.
\end{eqnarray}
Solving the three coupled equations above numerically   we can obtain a series of
$\sigma,\Phi,\overline{\Phi}$ at different temperature and chemical
potential.   The thermodynamical quantities  such as the
pressure,the quark number density,the entropy and the energy
density  can be computed with the thermodynamic relations:
\begin{eqnarray}
\nonumber P&=&-\frac{\Omega}{V};  \rho_{q}=-(\frac{\partial\Omega}{\partial\mu})_{T};\\
\nonumber S&=&-(\frac{\partial P}{\partial T})_{\mu};
 \varepsilon=TS+\mu\rho_{q}-P.
\end{eqnarray}
To this end, we can calculate the bulk viscosity from  Eq.(\ref{7}).

In this work we consider two-flavor quark matter. For numerical calculations, we choose the   parameters
 as followings\cite{r10}:  the global cutoff $\Lambda=0.651$
GeV, the quark current mass m=0.0055 GeV, the coupling constant
$g=5.04$GeV.  We also choose $T_{0}=0.27$GeV,
 the zero temperature quark condensation $|\sigma_{0}|=0.251^{3}$GeV  and $\omega_{0}=1$Gev. The vacuum energy density
$|\epsilon_{v}|^{1/4}=0.25$GeV

The temperature dependences of the order parameters for  chiral phase transition and de-confinement  phase transition $\sigma/\sigma_{0},\overline{\Phi},\Phi$  are plotted  in Fig.(\ref{fig1} ). It shows that the chiral phase transition temperature is
about $0.24$GeV with a quark chemical potential $\mu=0.2$GeV. This phase transition is a cross over. While the deconfinement phase transition might happen at higher temperature, although the Polyakov loops are not exact order paremeters for deconfinement  phase transition of QCD with quarks included.

\begin{figure}
\includegraphics[height=2.5in, width=3.5in]{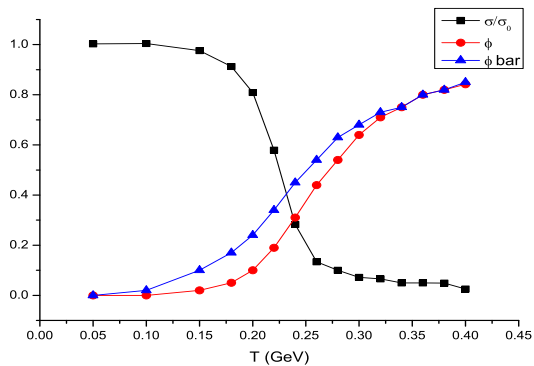}
\caption{\label{fig:epsart} Scaled chiral condensate and Polyakov loops as
functions of temperature at $\mu=0.2$GeV.}
\label{fig1}
\end{figure}

The numerical results for bulk viscosity are depicted in Fig.(\ref{fig2}) at different quark chemical potentials. One can see that the bulk viscosity has a sharp peak around the chiral phase transition temperature,just as the results of Masashi
Mizutani\cite{r9} .   It indicates that the finite  quark chemical potential increases the bulk viscosity with the same temperature.

We also computed the specific bulk viscosity, the ratio  of the bulk viscosity and entropy density , at finite temperature and density shown  in Fig.(\ref{fig3}). We show that this ratio starts to increase rapidly and blows up around the critical temperature. The result is in
agreement with the lattice results\cite{bulk1}

 The finite quark chemical potential  decreases the specific bulk viscosity though increases the bulk viscosity. This is because the finite chemical potential enhances the entropy density more rapidly than the bulk viscosity.
\begin{figure}
\includegraphics[height=2.5in, width=3.5in]{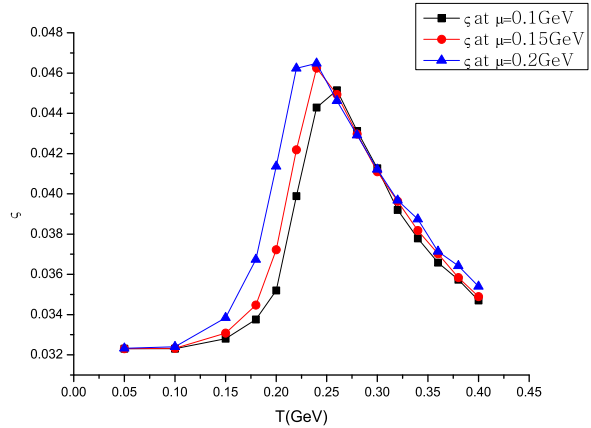}
\caption{\label{fig:epsart} Bulk viscosity at different chemical
potential and the increasing chemical potential raise the bulk
viscosity.}
\label{fig2}
\end{figure}

\begin{figure}
\includegraphics[height=2.5in, width=3.5in]{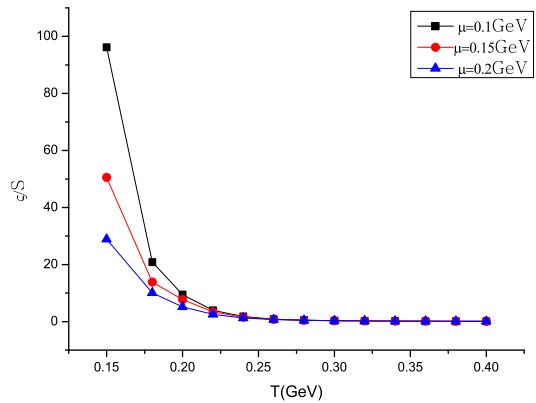}
\caption{\label{fig:epsart} The ratio of bulk viscosity to entropy
at different chemical potential.}
\label{fig3}
\end{figure}

In summary, We studied the bulk viscosity of hot quark matter at finite temperature and density within PNJL model by making use of the  the Kubo formula  and the  QCD low energy theorem.    We show that the bulk
viscosity has a sharp peak near the chiral phase transition, and the ratio
of bulk viscosity and the entropy  density  rises dramatically
in the vicinity of the  chiral   phase transition. These results agrees with that from lattice and other model calculations.  In addition, we show that  the increase of chemical potential raises the bulk viscosity  but decreases the ratio of the bulk viscosity and entropy density.

\begin{acknowledgments}
We would like to extend our gratitude to  Hai-chang Ren
 for helpful discussions.  This work is  supported partly by NSFC under grant Nos. 11135011, 11221504 and 10947002.
\end{acknowledgments}

\end{document}